\documentclass[twocolumn]{aastex7}

\newenvironment{Data Availability}{
    \section*{Data Availability}
    \begin{quote}
}{\end{quote}}
\usepackage{amsmath}
\begin{document}

\title{The relation between helium white dwarf mass and orbital period under two types of opacity}

\author[gname=Jian,sname=Mou]{Jian Mou} 
\affiliation{Yunnan Observatories, Chinese Academy of Sciences (CAS), Kunming 650216, People’s Republic of China}
\affiliation{School of Astronomy and Space Science, University of Chinese Academy of Sciences (UCAS), Beijing 100049, People’s Republic of China}
\email[show]{moujian@ynao.ac.cn}

\author[gname=Hai-Liang,sname=Chen]{Hai-Liang Chen} 
\affiliation{Yunnan Observatories, Chinese Academy of Sciences (CAS), Kunming 650216, People’s Republic of China}
\affiliation{International Centre of Supernovae, Yunnan Key Laboratory, Kunming 650216, P. R. China}
\email[show]{chenhl@ynao.ac.cn}

\author{Dengkai Jiang}
\affiliation{Yunnan Observatories, Chinese Academy of Sciences (CAS), Kunming 650216, People’s Republic of China}
\affiliation{International Centre of Supernovae, Yunnan Key Laboratory, Kunming 650216, P. R. China}
\email{}

\author{Hongwei Ge}
\affiliation{Yunnan Observatories, Chinese Academy of Sciences (CAS), Kunming 650216, People’s Republic of China}
\affiliation{International Centre of Supernovae, Yunnan Key Laboratory, Kunming 650216, P. R. China}
\email{}

\author{Lifu Zhang} 
\affiliation{Yunnan Observatories, Chinese Academy of Sciences (CAS), Kunming 650216, People’s Republic of China}
\affiliation{School of Astronomy and Space Science, University of Chinese Academy of Sciences (UCAS), Beijing 100049, People’s Republic of China}
\email{}

\author{Rizhong Zheng} 
\affiliation{Yunnan Observatories, Chinese Academy of Sciences (CAS), Kunming 650216, People’s Republic of China}
\affiliation{School of Astronomy and Space Science, University of Chinese Academy of Sciences (UCAS), Beijing 100049, People’s Republic of China}
\email{}

\author{Xuefei Chen}
\affiliation{Yunnan Observatories, Chinese Academy of Sciences (CAS), Kunming 650216, People’s Republic of China}
\affiliation{International Centre of Supernovae, Yunnan Key Laboratory, Kunming 650216, P. R. China}
\email{}

\author{Zhanwen Han}
\affiliation{Yunnan Observatories, Chinese Academy of Sciences (CAS), Kunming 650216, People’s Republic of China}
\affiliation{International Centre of Supernovae, Yunnan Key Laboratory, Kunming 650216, P. R. China}
\email{}



\begin{abstract}
Helium white dwarfs (He WDs) are end products of low-mass red giant donors in close binary systems via stable mass transfer or common envelope evolution. 
At the end of stable mass transfer, there is a well-known relation between the He WD mass and orbital period.  Although this relation has been widely investigated, the influence of different types of opacity at low temperatures is ignored. In this work, we modeled the evolution of WD binaries with stellar evolution code \textsf{MESA} and two types of opacity at low temperatures from Ferguson et al. (2005) and Freedman et al. (2008, 2014).  We investigated the relation between the WD mass and orbital period and compared these results with observations. We find that the relation derived from the opacity of Freedman et al. (2008, 2014) is below that from the opacity of Ferguson et al. (2005) and the relation derived from the opacity of Freedman et al. (2008, 2014) can better explain the observations.
In addition, we provided fitting formulae for the relations derived from the opacity of Freedman et al. (2008,2014) at different metallicities.

\end{abstract}



\section{introduction} 


All single low- and intermediate-mass stars eventually evolve into white dwarfs, which represent their final evolutionary stage. Due to their electron-degenerate stellar structure, they play a crucial role in our understanding of stellar formation and evolution, the history of the Galaxy, stellar populations, and the evolution of planetary systems \citep[e.g.][]{Wood1992,Debes2002,Althaus2010}. However, helium white dwarfs (He WDs) are widely recognized as products of binary evolution in close binary systems. This is because a single star can not evolve into a He WD within the age of the Universe. 
Stars with masses $\lesssim 2.3\ M_\odot$ can lose their envelope during the red giant branch (RGB) through stable Roche-lobe overflow (RLOF) or a common envelope (CE) phase, eventually leaving He WDs behind as remnants of their evolution. 
During the RGB, a star exhibits a characteristic core mass–stellar radius relation that is nearly independent of the mass of its hydrogen-rich envelope \citep{rw71,wrs83}. If the star undergoes stable RLOF, mass transfer occurs when the donor’s radius approaches its Roche-lobe radius ($R_{\mathrm{L}}$), which depends on the orbital separation and the mass ratio of the binary system \citep[][]{Paczy1971,Eggleton1983}. Consequently, the orbital period ($P_{\mathrm{orb}}$) at the end of mass transfer becomes closely related to the helium core mass ($M_{\mathrm{WD}}$) of the donor, establishing a well-defined relation \citep[][]{savo87,Joss1987,Rappaport1995,Tauris1999}. This correlation gives rise to the so-called mass–orbital period ($M_{\mathrm{WD}}$\textendash$P_{\mathrm{orb}}$) relation.

In earlier studies, \citet{Joss1987} first suggested that stable mass transfer in low-mass X-ray binaries (LMXBs) could lead to a correlation between the mass of the white dwarf and the final orbital period. Later, \citet{Rappaport1995} systematically investigated the core mass and radius of red giant stars at different metallicities and derived the $M_{\mathrm{WD}}$\textendash$P_{\mathrm{orb}}$ relation. They find that this relation is affected by the mixing length and also the metallicities. 
With the detailed calculation of binary evolution, \citet{Tauris1999} confirmed the results of \citet{Rappaport1995}, further showing that the mass transfer efficiency, the mode of mass loss, and magnetic braking have only minor effects on the relationship. 
Since then, the relation between WD mass and orbital period has been studied by different groups with different stellar evolution codes for different types of binaries \citep[e.g.][]{DeVito2010,Lin2011,Chenxf2013,Jia2014,Istrate2016}. But the $M_{\mathrm{WD}}$\textendash$P_{\mathrm{orb}}$ relations presented in these studies are very similar. 

Recently \citet{zcch21} have studied the influence of mass transfer schemes on the $M_{\mathrm{WD}}$\textendash$P_{\mathrm{orb}}$ relation. They found that the mass-orbital period relation from models with mass transfer scheme from \citet{kr90} is above the relation from models with mass transfer scheme from \citet{hte00}.

In addition, \citet{GaoShi-Jie2023} have investigated the influence of tidally enhanced stellar wind on the relation between the WD mass and orbital period. They found that strong winds can prematurely strip the envelopes of red giants before they fill their Roche lobes, suppressing core growth and significantly modifying the $M_{\mathrm{WD}}$\textendash$P_{\mathrm{orb}}$ relation. 

Although the relation $M_{\mathrm{WD}}$\textendash$P_{\mathrm{orb}}$ relation has been widely investigated, the influence of opacity at low temperature on this relationship has not been investigated. From the point of view of stellar evolution, the opacity will definitely influence the evolution of stars and hence the $M_{\mathrm{WD}}$\textendash$P_{\mathrm{orb}}$ relation as well. The purpose of this paper is to quantitatively study the impact of opacity at low temperature on the $M_{\mathrm{WD}}$\textendash$P_{\mathrm{orb}}$ relation. The structure of this paper is as follows. In Section~\ref{sec:2}, we describe the methods and physical assumptions adopted in our simulation. In Section~\ref{sec:3}, we present our theoretical evolutionary results. In Section~\ref{sec:dis}, we have a brief discussion and compare our results with observations. Finally, in Section~\ref{sec:con}, we have our conclusions.

\section{Method and Assumptions} \label{sec:2}
In our work, we use the one-dimensional stellar evolution code \textsc{MESA} \citep[MESA version r24.08.1,][]{Paxton2011,Paxton2013,Paxton2015,Paxton2018,Paxton2019}, to compute the evolution of CO WD+MS binary systems.
The initial mass of the main sequence donor ($M_\mathrm{{d,i}}$) ranges from 1.0 to 1.5 $M_\odot$ with a step of $0.10\;M_{\odot}$, while the initial white dwarf mass ($M_\mathrm{{CO,i}}$) ranges from 0.5 to 1.3 $M_\odot$ with a step of 0.1 $M_\odot$. The initial orbital period ($P_\mathrm{{orb,i}}$) spans from 0.5 to 8.0 days with a step of 0.1 days, from 8.0 to 30.0 with a step of 0.5 days, and from 30.0 to 530.0 days with a step of 1.0 days. 

In our models, the WD is treated as a point mass and we do not follow the evolution of its structure. For the initial chemical composition of donor stars, we adopt three different kinds of chemical compositions. Specifically, we consider three different initial metallicities, that is, $Z = 0.02,\; 0.001,\; 0.0001$. The initial hydrogen mass fraction is computed with $X = 0.76-3Z$ and the initial helium abundance $Y = 1.0-X-Z$. We set the mixing length parameter to be 2.0 and ignore overshooting in our calculation. 

Regarding the low-temperature radiative opacity, the default choice in MESA is the table of \citet{Ferguson2005}, which includes contributions from molecules and condensates (grains) and extends down to temperatures of $\approx 500\,\mathrm{K}$. Another low-temperature opacity table of interest is that of \citet{Freedman2008, Freedman2014}. 
This table includes molecular opacities but neglects condensates, and reaches down to $\approx 75\,\mathrm{K}$. We perform our calculations with both opacity prescriptions.

The atmospheric boundary condition for the donor star is computed using the Eddington grey relation $T^4(\tau) = \frac{3}{4} T_{\mathrm{eff}}^4 \left( \tau + \frac{2}{3} \right)$, with a fixed opacity that matches the local temperature and pressure throughout the atmosphere. The equation of state is a combination of PC \citep[][]{Potekhin2010}, FreeEOS \citep[][]{Irwin2012}, OPAL \citep[][]{Rogers2002}, SCVH \citep[][]{Saumon1995}, HELM \citep[][]{Timmes2000} and Skye \citep[][]{Jermyn2021}. We also adopt the nuclear network \texttt{cno\_extras.net} for our calculation.

In addition, we take into account the effect of the stellar wind using the prescription of \citet{Reimers1975},

\begin{equation}
\dot{M}_{\mathrm{d},\mathrm{wind}} = -4 \times 10^{-13} \, M_{\odot} \, \mathrm{yr}^{-1} \, \eta \left( \frac{R_\mathrm{d}}{R_{\odot}} \right) \left( \frac{L_\mathrm{d}}{L_{\odot}} \right) \left( \frac{M_{\odot}}{M_\mathrm{d}} \right),
\end{equation}
where $\eta$ equal 0.5, and $R_\mathrm{d}$ and $L_\mathrm{d}$ are the radius and luminosity of the donor star, respectively.

We adopt the binary mass transfer scheme of \citet{Ritter1988} to compute the mass transfer rate,
\begin{equation}
\dot{M}_{\rm RLOF} \propto \frac{R_{\mathrm{RL},\mathrm{d}}^3}{M_\mathrm{d}} \exp \left( \frac{R_\mathrm{d} - R_{\mathrm{RL},\mathrm{d}}}{H_\mathrm{p}} \right),
\end{equation}
where $R_\mathrm{d}$ and $R_{\mathrm{RL,d}}$ are the radius of the donor and the radius of the Roche lobe, respectively; $H_\mathrm{p}$ is the pressure scale height.

Regarding the angular momentum loss, we consider three different mechanisms, i.e, magnetic braking (MB), gravitational wave radiation (GWR), and mass loss (ML).

MB is an very important angular momentum loss mechanism in binary evolution and still very uncertain \citep[e.g.][]{Chenhl2022}. From previous studies \citep[e.g.][]{Tauris1999}, we know that the angular momentum loss will not influence the $M_{\rm WD}-P_{\rm orb}$ relation. Therefore, we adopt the prescription from  \citet{Rappaport1983}, which is the default implementation in MESA,

\begin{equation}
\dot{J}_{\mathrm{mb}} = -3.8 \times 10^{-30} \, M_{\mathrm{d}} R_\odot^4 \left( \frac{R_{\mathrm{d}}}{R_\odot} \right)^{\gamma_{\mathrm{mb}}} \Omega^3 \, \mathrm{dyn\,cm}, 
\end{equation}
where $M_{\mathrm{d}}$ is the mass of the donor, $R_{\mathrm{d}}$ is the radius of the donor, $\Omega$ is the orbital angular velocity and $\gamma_{\mathrm{mb}}$ is set to be 4.0 in our simulations.

Given that no strong magnetic fields form in a donor with a very small convective envelope, the magnetic braking should be weaker in this scenario.
Following \citet{Podsiadlowski2002}, if the mass of the convective envelope falls below 2\% of the total donor mass, we implement this effect by applying a scaling factor of \( e^{1 - 0.02 / q_\mathrm{conv,env}} \), where \( q_\mathrm{conv,env} \) is the mass fraction of the donor's convective envelope. For the onset of MB, we adopt the default criteria used in this version of MESA: the mass fraction of the convective envelope must satisfy $10^{-6} \leq q_{\mathrm{conv,env}} < 0.99$, while the mass fraction of the convective core must satisfy $q_{\mathrm{conv,core}} \leq 0.01$.

When the orbital period of a binary system decreases to just a few hours, the GWR plays a crucial role in the orbital evolution. Here, we adopt the formulation from \citet{Landau1975}, 

\begin{equation}
\dot{J}_\mathrm{gr} = -\frac{32}{5c^5} \left( \frac{2\pi G}{P_\mathrm{orb}} \right)^{7/3} \frac{(M_\mathrm{d} M_\mathrm{CO})^2}{(M_\mathrm{d} + M_\mathrm{CO})^{2/3}},
\end{equation}
where \( G \) is the gravitational constant, \( c \) is the speed of light in vacuum, $P_{\rm orb}$ is the binary orbital period, \( M_\mathrm{d} \) is the mass of the donor star, and \( M_\mathrm{CO} \) is the mass of the CO WD. 

The loss of angular momentum due to mass loss is computed with the following equation,

\begin{equation}
\centering
\dot{J}_{\mathrm{ml}} = 
   (\dot{M}_{\mathrm{d},\mathrm{wind}}M_{\mathrm{CO}}^2
+(1-\eta)\dot{M}_{\mathrm{RLOF}} M_\mathrm{d}^2)
 \times \frac{a^2\Omega}{(M_\mathrm{d}+M_\mathrm{CO})^2},
\end{equation}

where $\dot{M}_{\mathrm{d},\mathrm{wind}}$ is the stellar wind of the donor star, $\dot{M}_{\mathrm{RLOF}}$ is the mass transfer rate due to Roche-lobe overflow, $a$ is the binary separation, $\Omega$ is the orbital angular velocity, and $\eta$ is the total accumulation efficiency. Since $\dot{M}_{\mathrm{d},\mathrm{wind}}$ and $\dot{M}_{\mathrm{RLOF}}$ are defined as negative values in MESA, we do not introduce an extra minus sign in this equation.

The mass accumulation efficiency of CO WD is strongly dependent on its mass and accretion rate \citep{yps05,cwyp+19}. Here we adopt the prescription from \citet{hkn99} for the accretion of H-rich material. The specific prescription is given below. When $|\dot{M}_\mathrm{RLOF}| > \dot{M}_\mathrm{cr}$, the accreted H will burn stably on the surface of WD at a rate of $\dot{M}_\mathrm{cr}$ and the excess mass will be lost in the form of an optically thick wind. If $\frac{1}{2} \dot{M}_\mathrm{cr} \leq |\dot{M}_\mathrm{RLOF}| < \dot{M}_\mathrm{cr}$, all the accreted material burns stably and no mass is lost. If $\frac{1}{8} \dot{M}_\mathrm{cr} \leq |\dot{M}_\mathrm{RLOF}| < \frac{1}{2}\dot{M}_\mathrm{cr}$, hydrogen burning is unstable and assuming no mass is lost due to the weak H-shell flash. If  $|\dot{M}_\mathrm{RLOF}| < \frac{1}{8} \dot{M}_\mathrm{cr}$, the H-shell flash is so strong and no mass is retained.  The critical accretion rate is computed with the following formula, 
\begin{equation}
\dot{M}_{\mathrm{cr}} = 5.3 \times 10^{-7}\;M_{\odot}/{\rm yr} \frac{(1.7 - X)}{X} (M_{\mathrm{CO}} - 0.4),
\end{equation}
where $X$ is the surface hydrogen abundance of the donor and $M_{\mathrm{CO}}$ is the mass of the accretor in an unit of solar mass.

Therefore the mass growth rate of helium layer of the accretor $\dot{M}_\mathrm{He} = \eta_\mathrm{H} |\dot{M}_\mathrm{RLOF}|$, 
where 

\begin{equation}
\eta_{\mathrm{H}} =
\left\{
\begin{array}{ll}
\dot{M}_{\mathrm{cr}} / |\dot{M}_{\mathrm{RLOF}}|, & \text{if } |\dot{M}_{\mathrm{RLOF}}| > \dot{M}_{\mathrm{cr}}; \\
1, & \text{if } \dot{M}_{\mathrm{cr}} \geq |\dot{M}_{\mathrm{RLOF}}| \geq \frac{1}{8} \dot{M}_{\mathrm{cr}}; \\
0, & \text{if } |\dot{M}_{\mathrm{RLOF}}| < \frac{1}{8} \dot{M}_{\mathrm{cr}}.
\end{array}
\right.
\end{equation}

As the He accumulates on the surface of WD, it will burns stably or unstably into CO, which also depends on the WD mass and the mass growth rate of He. Here we adopt the prescription from \citet{Kato2004} to compute the mass growth rate for He burning, $\eta_{\rm He}$. Therefore, the total mass growth rate of the WD $\dot{M}_\mathrm{CO} = \eta_\mathrm{He} \dot{M}_\mathrm{He} = \eta_\mathrm{He}\eta_{\rm H} |\dot{M}_\mathrm{RLOF}|$ and $\eta = \eta_{\rm He}\eta_{\rm H}$.

In our simulation, we limit the maximum evolutionary age of our models to the age of the Universe, 13.7 Gyr. In addition, we exclude these binary systems with maximum mass transfer rate larger than $10^{-4}\;M_{\odot}/{\rm yr}$, which are assumed to enter common envelope evolution.

\section{results} \label{sec:3}

\subsection{Examples of binary evolution}

\begin{figure*}[htbp]
  \centering
  \includegraphics[width=\linewidth]{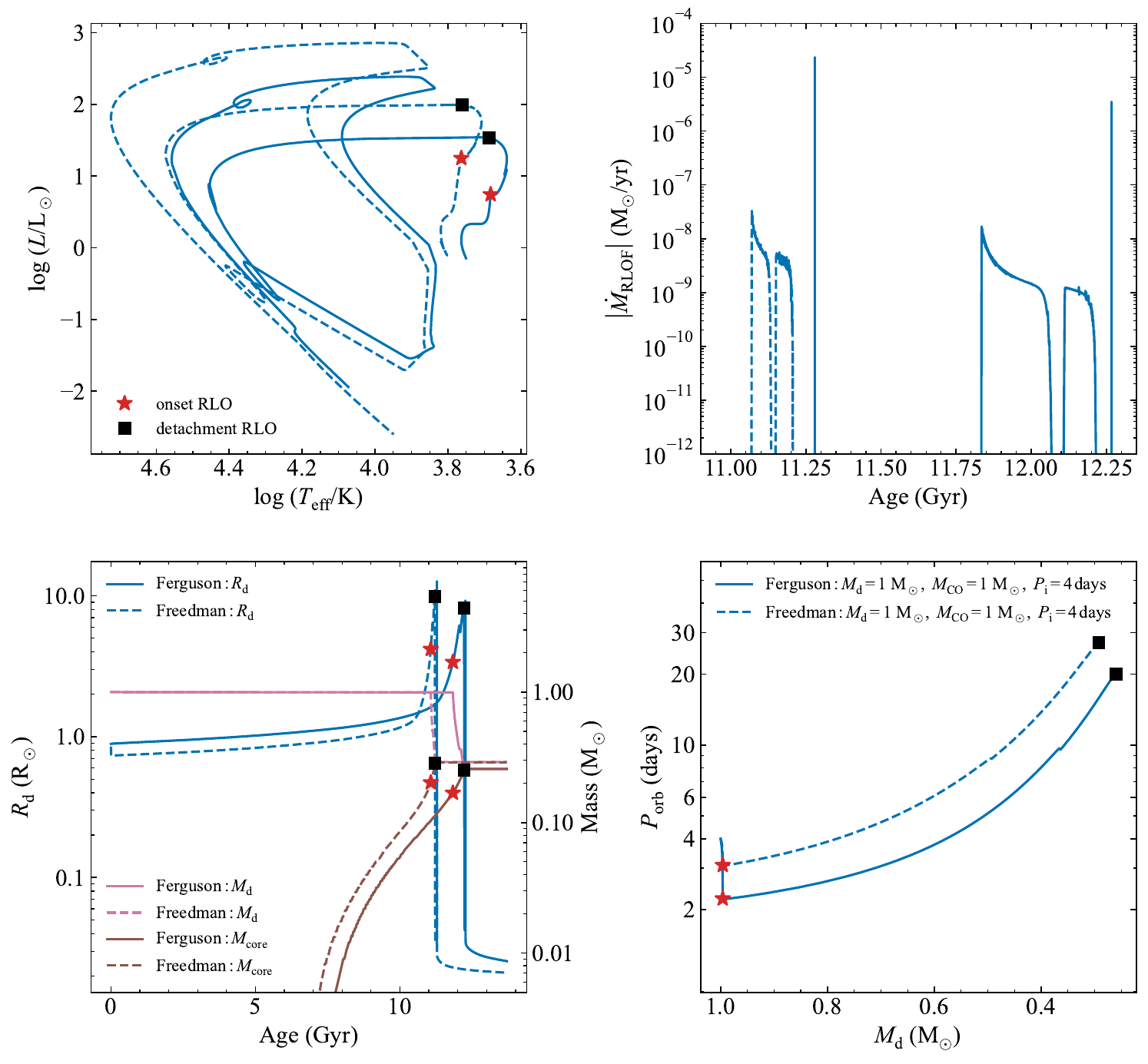}
  \caption{Evolution of a binary system with an initial donor mass of $1\,M_\odot$, an initial carbon--oxygen white dwarf mass of $1\,M_\odot$, and an initial orbital period of $4\,\mathrm{days}$, calculated with two different low-temperature radiative opacity tables, i.e., \citet{Ferguson2005} and \citet{Freedman2008,Freedman2014}. Top left panel: Hertzsprung--Russell diagram; Top right panel: Evolution of mass-transfer rate as a function of age; Bottom left panel: Evolution of the radius (blue color), stellar mass (pink color) and He core mass (brown color) as a function of age. Bottom right panel: Evolution of orbital period as a function of donor mass. In all the panels, the solid lines are for the models with the opacity of \citet{Ferguson2005} and the dashed lines are for the models with the opacity of \citet{Freedman2008,Freedman2014}. The red stars denote the onset of the mass-transfer phase, while the black squares indicate the end of the second mass transfer phase. In the upper right panel, the very short mass transfer phases at the end of evolution are due to the H-shell flashes (see the HR diagram).
  }
  \label{fig:track}
\end{figure*}

In Fig.~\ref{fig:track}, we show the evolution of a binary system with two types of opacity. The initial binary parameters in this example are $M_{\rm CO} = 1.0\;M_{\odot}$, $M_{\rm d} = 1.0\;M_{\odot}$ and ${P_{\rm orb} = 4.0\;}$days. In the two cases, the donor stars start mass transfer on the RGB and the systems evolve into double WD systems eventually. In the model with the opacity of \citet{Freedman2008,Freedman2014}, the donor star evolves faster and starts mass transfer earlier compared with the model with the opacity of \citet{Ferguson2005}. During the RGB, at the point where the core mass is the same, the radius of the donor star is smaller in the model with the opacity of \citet{Freedman2008,Freedman2014}. 
Therefore, the core mass is larger at the onset of mass transfer and also at the end of mass transfer in the model with the opacity of \citet{Freedman2008,Freedman2014} . The orbital period decreases firstly because of the effect of magnetic braking and increases later due to mass transfer.  At the end of mass transfer, the orbital period is larger in the model with the opacity of \citet{Freedman2008,Freedman2014} compared with that in the model with the opacity of \citet{Ferguson2005}. 
Consequently the He WD mass at the end of mass transfer is larger  and the final orbital period is larger as well in the model with \citet{Freedman2008,Freedman2014}. In both cases, there is a very short but high-rate of mass transfer phase at their end of evolution, which is due to the H-shell flash.

\subsection{Mass–Orbital Period ($M_{\mathrm{WD}}$\textendash$P_{\mathrm{orb}}$) Relation}

\begin{figure*}[htb]
  \centering
  \includegraphics[width=\linewidth]{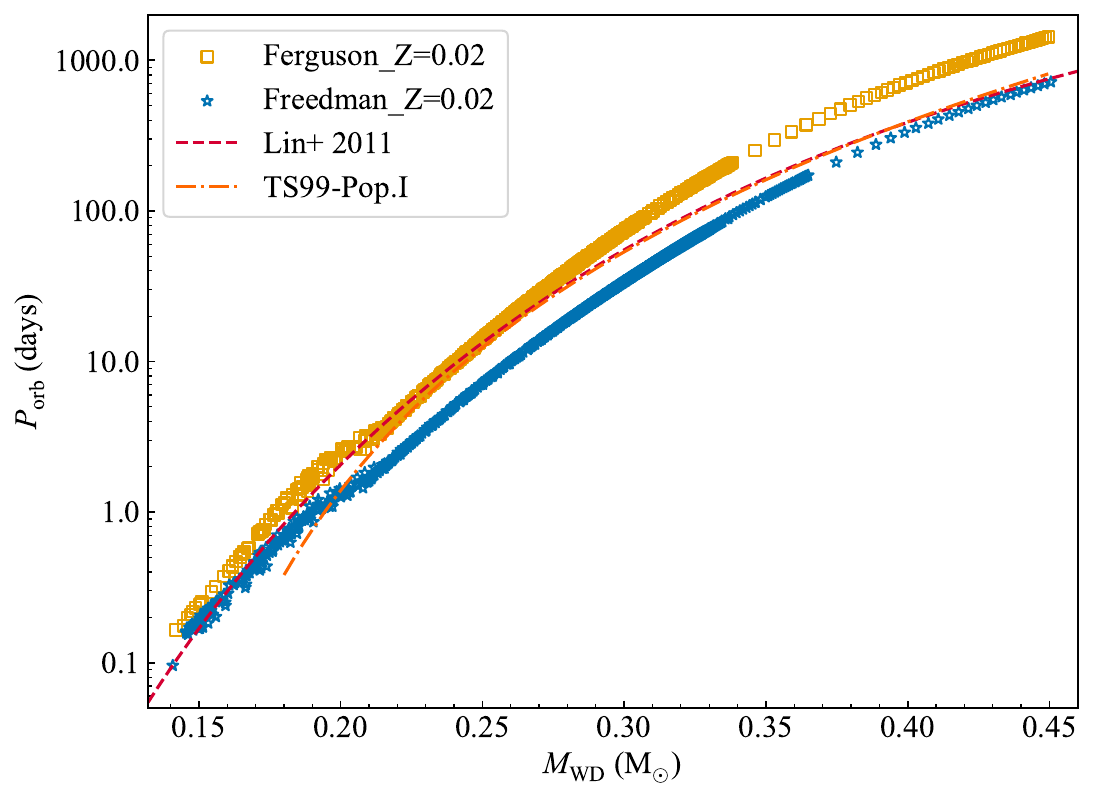}
  \caption{$M_{\mathrm{WD}}$\textendash$P_{\mathrm{orb}}$ relations obtained with binary models with different opacity tables at solar metallicity. Here the orbital period and WD mass represents the values at the end of mass transfer. The yellow squares are for the low-temperature opacity table from \citet{Ferguson2005}, while the blue pentagrams are for the opacity tables from \citet{Freedman2008,Freedman2014}. The red dashed line shows the $M_{\mathrm{WD}}$\textendash$P_{\mathrm{orb}}$ relation for solar metallicity from \citet{Lin2011}, and the yellow dash-dotted line denotes the relation for population I given by \citet{Tauris1999}.}
  \label{fig:mass-period-1}
\end{figure*}

Based on the initial evolutionary grids described in Section~\ref{sec:2}, we obtained the relations $M_{\mathrm{WD}}$\textendash$P_{\mathrm{orb}}$ under the low-temperature radiative opacity of \citet{Ferguson2005} and \citet{Freedman2008,Freedman2014}. As shown in Fig.~\ref{fig:mass-period-1}, the $M_{\mathrm{WD}}$\textendash$P_{\mathrm{orb}}$ relation computed using the \citet{Freedman2008,Freedman2014} opacity tables lies systematically below that obtained with the \citet{Ferguson2005} opacity. Specifically, for the same helium WD mass, the \citet{Freedman2008,Freedman2014} opacity results in shorter orbital periods.  This difference is thought to result from differences in the stellar radius during the RGB, as indicated in the previous section.

Compared with the $M_{\rm WD}-P_{\rm orb}$ relation from other studies, the relation obtained with the opacity from \citet{Ferguson2005} at low WD mass part is consistent with the relation from \citet{Lin2011} and is above the relation from \citet{Lin2011} and \citet{Tauris1999} at the high WD mass part. 
For the relation obtained with the opacity from \citet{Freedman2008, Freedman2014}, it is consistent with the relation from \citet{Lin2011} at low and high mass end. But it is below their relation in the intermediate WD mass part. 

\subsection{$M_{\rm WD}-P_{\rm orb}$ relations at different metallicities}

\begin{figure*}[htbp]
  \centering
  \includegraphics[width=\linewidth]{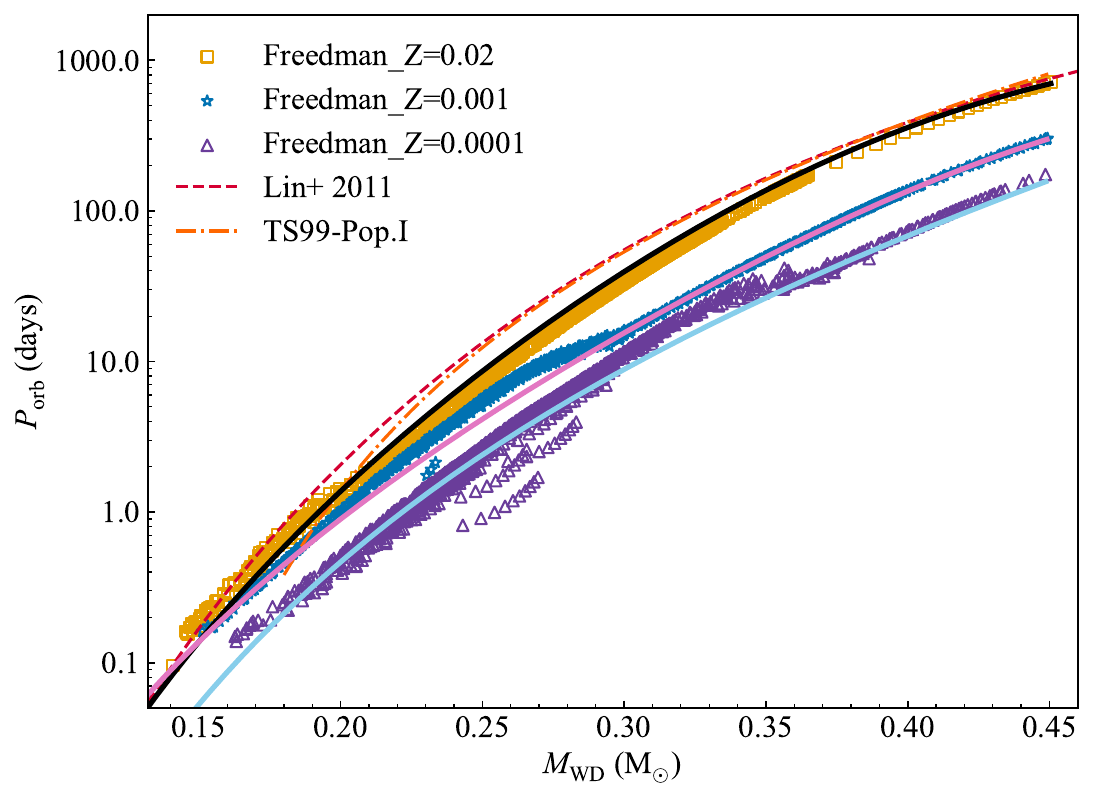}
  \caption{The $M_{\rm WD}-P_{\rm orb}$ relations computed with the opacity of \citet{Freedman2008,Freedman2014}. The yellow squares, blue pentagons, and purple upward triangles represent the theoretical results for metallicities of Z = 0.02, 0.001, and 0.0001, respectively. The black, pink, and light blue solid lines correspond to the fitted relations for each of these metallicities.}
  \label{fig:fit-mass-period}
\end{figure*}

\begin{table}[htbp]
\centering
\caption{Best-fitting Parameters in Eq.~\ref{eq:fit} for the $M_{\mathrm{WD}}$\textendash$P_{\mathrm{orb}}$ Relations at different metallicities.}
\label{tab:fit_params_1}
\begin{tabular}{lccccc}
\hline
metallicity & $\mathrm{a}$ & $\mathrm{b}$ & $\mathrm{c}$ & $\mathrm{d}$ & $\mathrm{e}$ \\
\hline
Z=0.02 & $1.00 \times 10^5$ & 7.50 & 0.70 & $-4.2$ & 16.00 \\
Z=0.001 & $1.05 \times 10^5$ & 6.00 & 4.80 & $-27$ & 65.70 \\
Z=0.0001 & $1.15 \times 10^5$ & 8.25 & 0.40 & 4.35 & $-7.0$ \\
\hline
\end{tabular}
\end{table}

In Fig.~\ref{fig:fit-mass-period}, we present the $M_{\mathrm{WD}}$\textendash$P_{\mathrm{orb}}$ relations computed using the \citet{Freedman2008,Freedman2014} low-temperature radiative opacity tables at three different metallicities. The solid lines in different colors represent the fitted relations for different cases. 
Similar to \citet{Tauris1999}, we can find that the relation at lower metallicity is below the relation at higher metallicity. 
This is mainly because the radii of a RG star with a same total mass and core mass at lower metallicity is smaller than that at higher metallicity.
In addition, we have fitted this results with fitting functions from \citet{Lin2011},

\begin{equation}
\label{eq:fit}
P_\mathrm{orb} = \mathrm{a} M_\mathrm{WD}^\mathrm{b} \left( \mathrm{c} + \mathrm{d} M_\mathrm{WD}^2 + \mathrm{e} M_\mathrm{WD}^4 \right)^{-3/2}\ \mathrm{days},
\end{equation}
where $M_\mathrm{WD}$ is the mass of the helium white dwarf in units of solar mass. The best-fit parameters for each case are summarized in Table~\ref{tab:fit_params_1}.

\section{Discussion}
\label{sec:dis}
\subsection{Comparison with the observations of extremely low-mass white dwarfs binaries}

\begin{figure*}[htbp]
  \centering
  \includegraphics[width=\linewidth]{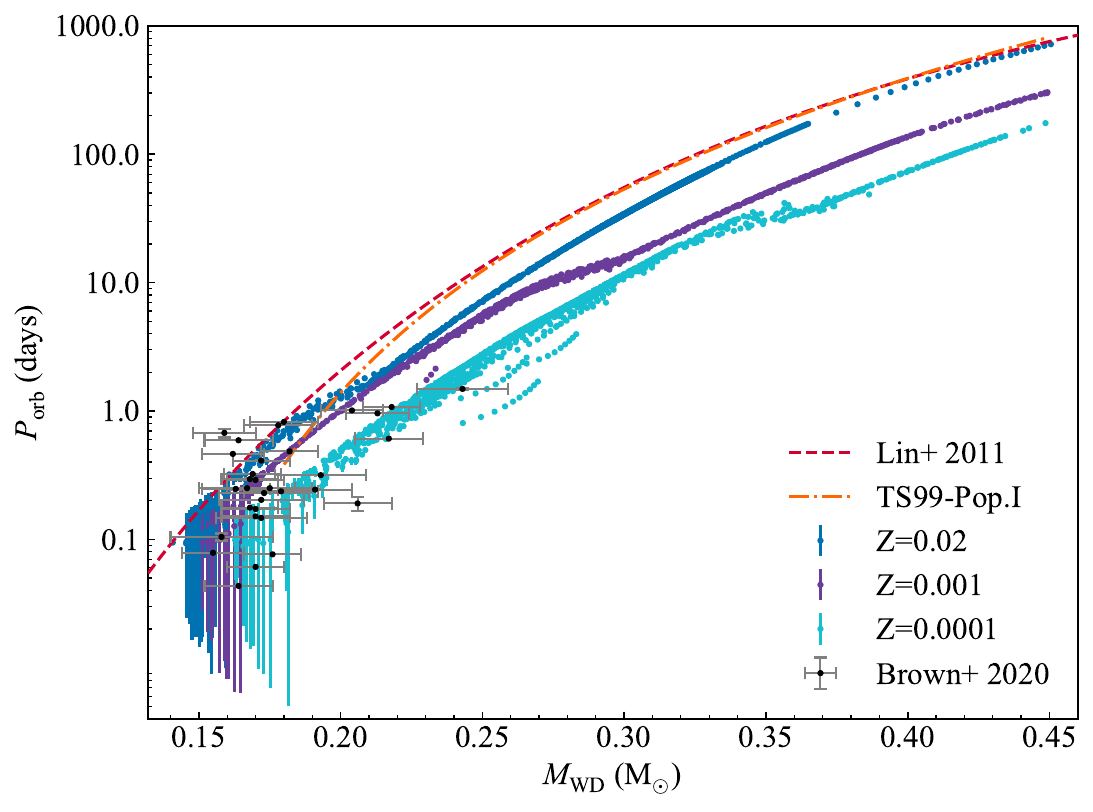}
  \caption{ Comparison of the $M_{\rm WD}-P_{\rm orb}$ relation derived from the models with the opacity of \citet{Freedman2008,Freedman2014} with the observations of extremely low-mass WD binaries. The dark blue, purple, and light blue symbols represent the theoretical results for metallicities Z = 0.02, 0.001 and 0.0001, respectively. At the low mass end, for a given WD mass, there is an orbital period range which is due to gravitational wave radiation. The upper limit corresponds to the orbital period at the end of mass transfer, while the lower limit is taken as the orbital period either at the onset of the mass transfer from He WD to the CO WD or at the Hubble time, whichever occurs first. The observational systems (i.e. black points with error bars) are selected from the \citet{Brown2020} sample using the method described in \citet{Lizw2019}, representing systems formed via stable RLOF.}
  \label{fig:obs1}
\end{figure*}

To assess the validity of our theoretical results at the low-mass end of the $M_{\mathrm{WD}}$\textendash$P_{\mathrm{orb}}$ relation, we compare them with the observations of extremely low mass white dwarf (ELM WD) binaries from the ELM Survey. 
In this work, we adopt the "clean sample" published by \citet{Brown2020}, which includes measurements of the binary orbital period, the ELM WD mass, and the minimum companion mass. 
Following the method proposed by \citet{Lizw2019} (see their Sec. 4.2.3), we classify the clean sample into two categories based on their evolutionary channels: systems from stable RLOF and CE channel. And we take these samples from stable RLOF for comparison.

In Fig.~\ref{fig:obs1}, we compare our theoretical results with the observations of ELM WD binaries. The dark blue, purple, and light blue symbols represent theoretical predictions at different metallicities. It is worth noting that there is an orbital period range for a given WD mass at the lower mass part. 
This is mainly because the GWR is strong for these systems with short orbital periods after mass transfer. 
And the He WD may fill its Roche lobe and transfer material to the CO WD \citep[e.g.][]{Tauris18,Chenhl2022}. The orbital period range corresponds to the detached phases of double WDs during their evolution. 
For each period range, the lower boundary corresponds to the onset of mass transfer from He WD to the CO WD or the evolutionary age equals to the Universe age. From this plot, we can find that our relations can well explain the observed ELM WD samples.

\subsection{Comparison with the observations of binary millisecond pulsars}

\begin{figure*}[htbp]
  \centering
  \includegraphics[width=\linewidth]{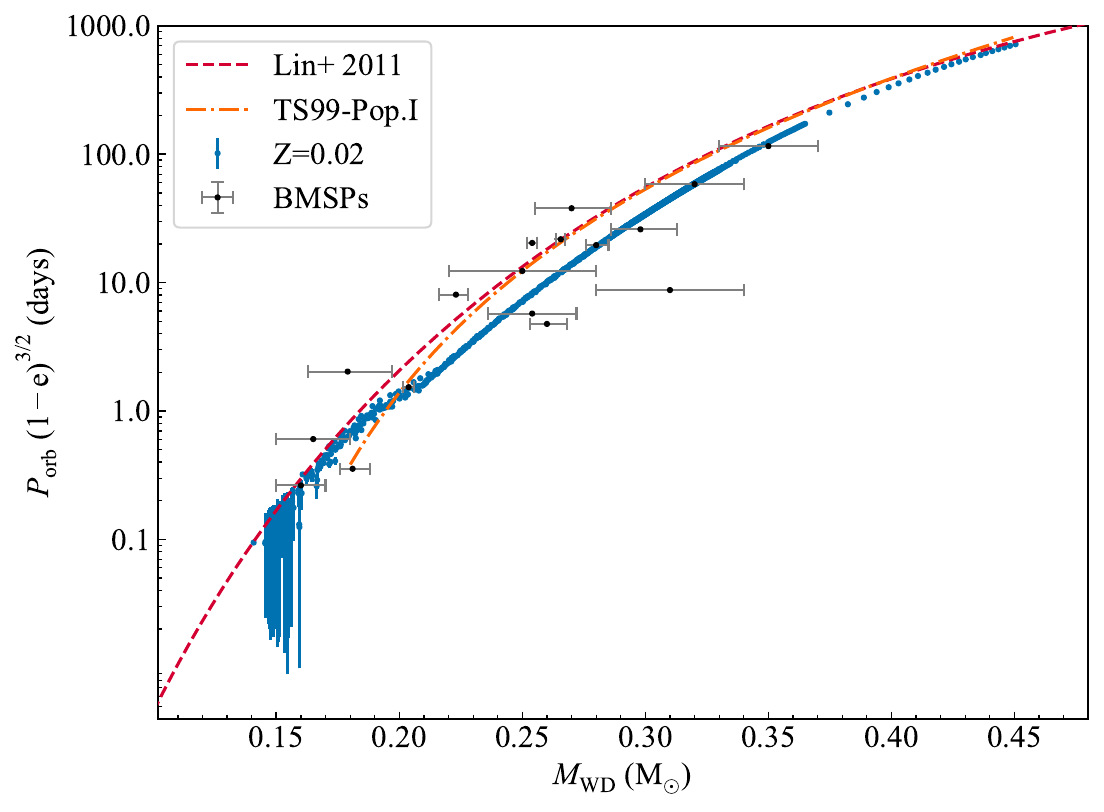}
  \caption{Comparison of the $M_{\rm WD}-P_{\rm orb}$ relation derived from models with \citet{Freedman2008,Freedman2014} opacity at solar metallicity with observations of BMSPs with He WD companions listed in Tab.~\ref{tab:hewd_bmsp}. The relations by \citet{Tauris1999} and \citet{Lin2011} are over-plotted.}
  \label{fig:obs2}
\end{figure*}

From previous studies, we know that the $M_{\rm WD}-P_{\rm orb}$ relation not only exits in double WD binaries but also in binary millisecond pulsars (BMSPs). In this section, we compare our results with the observations of millisecond pulsar + He WD binaries.

Currently there are 638 BMSPs with spin period $P_{\mathrm{spin}} \leq 30$ ms in the ATNF Pulsar Catalogue\footnote{\url{https://www.atnf.csiro.au/research/pulsar/psrcat/}} \citep{manchester2005}. 
Following \citet{Smedley2014}, we select these binary systems with both well measured pulsar masses and He WD masses. 
In addition, we exclude these systems with orbital period smaller than 1.0 days and located in globular cluster.  
Then we can find 17 BMSPs, whose parameters are listed in Table~\ref{tab:hewd_bmsp}.

In Fig.~\ref{fig:obs2}, we compare the $M_{\rm WD}-P_{\rm orb}$ relation at solar metallicity with these BMSPs listed in Tab.~\ref{tab:hewd_bmsp}. 
From this plot, we can find that our new relation can better explain these observations. 

\begin{table*}[htbp]
\centering
\caption{A sample of BMSPs with well measured pulsar and companion masses}
\begin{tabular}{lcccccc}
\hline
\hline
Name & $P_{\rm spin}$ (ms) & $P_{\rm orb}$ (days) & $M_{\rm p}/M_\odot$ & $M_{\rm comp}/M_\odot$ & eccentricity ($e$) & Reference \\
(1) & (2) & (3) & (4) & (5) & (6) & (7) \\
\hline
J0955$-$6150 & 1.99936049 & 24.578393 & $1.71 \pm 0.02$ & $0.254 \pm 0.002$ & 0.1175 & \citet{Serylak2022} \\
J1543$-$5149 & 2.05696039 & 8.0607731 & $1.349^{+0.043}_{-0.061}$ & $0.223^{+0.005}_{-0.007}$ & $2.133\times 10^{-5}$ & \citet{Chisabi2025}\\
J0218+4232 & 2.32309053 & 2.0288461 & $1.49^{+0.23}_{-0.20}$ & $0.179^{+0.018}_{-0.016}$ & $6.34\times 10^{-6}$ & \citet{Tan2024}\\
J1125$-$6014 & 2.63038078 & 8.7526036 & $1.5 \pm 0.2$ & $0.31 \pm 0.03$ & $6\times 10^{-6}$ & \citet{Reardon2021}\\
J0740+6620 & 2.88573641 & 4.7669446 & $2.14^{+0.10}_{-0.09}$ & $0.26^{+0.008}_{-0.007}$ & $5.1\times 10^{-6}$ & \citet{Cromartie2020}\\
J1909$-$3744 & 2.94710806 & 1.5334494 & $1.438 \pm 0.024$ & $0.2038 \pm 0.0022$ & $1.35\times 10^{-7}$ & \citet{Jacoby2005}\\
J1012$-$4235 & 3.10114124 & 37.972463 & $1.44^{+0.13}_{-0.12}$ & $0.27^{+0.016}_{-0.015}$ & $3.457\times 10^{-4}$ & \citet{Gautam2024}\\
J1946+3417 & 3.17013917 & 27.019947 & $1.828 \pm 0.022$ & $0.2656 \pm 0.0019$ & 0.1345 & \citet{Barr2017}\\
J0751+1807 & 3.47877083 & 0.2631442 & $1.64 \pm 0.15$ & $0.16 \pm 0.01$ & $3.3\times 10^{-6}$ & \citet{Desvignes2016}\\
J2234+0611 & 3.57658163 & 32.001401 & $1.353^{+0.014}_{-0.017}$ & $0.298^{+0.015}_{-0.012}$ & 0.1293 & \citet{Stovall2019}\\
J1853+1303 & 4.09179738 & 115.65378 & $1.4 \pm 0.7$ & $0.35 \pm 0.02$ & $2.36\times 10^{-5}$ & \citet{Gonzalez2011}\\
J1950+2414 & 4.30477548 & 22.191371 & $1.496 \pm 0.023$ & $0.280^{+0.005}_{-0.004}$ & 0.0798 & \citet{Zhu2019}\\
J1910+1256 & 4.98358401 & 58.466742 & $1.6 \pm 0.6$ & $0.32 \pm 0.02$ & $2.3\times 10^{-4}$ & \citet{Gonzalez2011}\\
J1012+5307 & 5.25574910 & 0.6046727 & $1.72 \pm 0.16$ & $0.165 \pm 0.015$ & 0 & \citet{Mata2020}\\
J1857+0943 & 5.36210054 & 12.327171 & $1.5 \pm 0.2$ & $0.25 \pm 0.03$ & $2.17\times 10^{-5}$ & \citet{Reardon2016}\\
J0437$-$4715 & 5.75745194 & 5.7410459 & $1.76 \pm 0.2$ & $0.254 \pm 0.018$ & $1.918\times 10^{-5}$ & \citet{Verbiest2008}\\
J1738+0333 & 5.85009586 & 0.3547907 & $1.47^{+0.07}_{-0.06}$ & $0.181^{+0.007}_{-0.005}$ & $3.5\times 10^{-7}$ & \citet{Antoniadis2012}\\
\hline
\end{tabular}

\tablecomments{(1)-Pulsar name.(2) Spin period. (3) Orbital period. (4) Pulsar mass. (5) He WD mass. (6) Eccentricity. (7) Reference for the binary parameters.}
\label{tab:hewd_bmsp}
\end{table*}

\section{Conclusion}
\label{sec:con}

With the stellar evolution code \textsc{mesa}, we computed the evolution of a grid of WD binaries at different metallicities and investigated the relation between the He WD mass and orbital period. For each metallicity, we adopted two types of low temperature opacity. One opacity is from \citet{Ferguson2005}, the other one is from \citet{Freedman2008,Freedman2014}. The maim conclusions are as follows.

\begin{enumerate}

\item The $M_{\rm WD}-P_{\rm orb}$ relation derived from the opacity of \citet{Freedman2008, Freedman2014} is below that from the opacity of \citet{Ferguson2005}.

\item We provide the $M_{\rm WD}-P_{\rm orb}$ relations derived from the opacity of \citet{Freedman2008, Freedman2014} at metallicity $Z = 0.02, 0.001, 0.0001$ and fitting formulae for them. 

\item The $M_{\rm WD}-P_{\rm orb}$ relation from the models with the opacity of \citet{Freedman2008,Freedman2014}
can well explain both the observations of ELM WD binaries and BMSPs with He WD binaries. 

\end{enumerate}

\begin{acknowledgments}
This work is partially supported by the National Natural Science Foundation of China (grant Nos. 12288102, 12090040/12090043, 12333008, 12422305, 12473033, 12525304 and 12125303), the National Key R$\&$D Program of China (grant Nos. 2021YFA1600400/401, 2021YFA1600403), the Strategic Priority Research Program of the Chinese Academy of Sciences (grant No. XDB1160201), the CAS "Light of West China", the Young Talent Project of Yunnan Revitalization Talent Support Program, the Yunnan Fundamental Research Project (No. 202401BC070007), the Yunnan Revitalization Talent Support Program$-$Science $\&$ Technology Champion Project (No. 202305AB35003), the International Centre of Supernovae (No. 202201BC070003), Yunnan Key Laboratory (No. 202302AN360001). The authors gratefully acknowledge the “PHOENIX Supercomputing Platform” jointly operated by the Binary Population Synthesis Group and the Stellar Astrophysics Group at Yunnan Observatories, CAS. We are grateful to the MESA council for the MESA instrument papers and website.
\end{acknowledgments}

\begin{Data Availability}
The inlist and src files from MESA required to reproduce our simulation are available at: \dataset[doi:10.5281/zenodo.17667065]{https://doi.org/10.5281/zenodo.17667065}.
\end{Data Availability}

\software{MESA \citep[MESA version r24.08.1,][]{Paxton2011,Paxton2013,Paxton2015,Paxton2018,Paxton2019}, 
Astropy \citep{Astropy2013,Astropy2018}, Numpy \citep{numpy2020}, Matplotlib \citep{Matplotlib2007}, Jupyter-Lab (https://jupyter.org)
          }




\bibliography{sample7}{}
\bibliographystyle{aasjournalv7}



\end{document}